\documentclass[preprint,times,3p,twocolumn]{elsarticle}  
\usepackage{latexsym}
\usepackage{amssymb}
\usepackage{amsmath}
\usepackage{graphicx}
\usepackage{multirow}

\journal{Physics Letters B}

\begin{document}

\begin{frontmatter}

\title{Did TOTEM experiment discover the Odderon?}

\author[em]{Evgenij Martynov}
\ead{martynov@bitp.kiev.ua}

\author[bn]{Basarab Nicolescu}
\ead{basarab.nicolescu@gmail.com}

\address[em]{Bogolyubov Institute for Theoretical Physics, Metrologichna 14b, Kiev, 03680 Ukraine}
\address[bn]{Faculty of European Studies, Babes-Bolyai University, Emmanuel de Martonne Street 1, 400090 Cluj-Napoca, Romania}

\begin{abstract}
The present study shows that the new TOTEM datum $\rho^{pp} = 0.098  \pm  0.01$     can be considered as the first experimental discovery of the Odderon, namely in its maximal form.
\end{abstract}


\end{frontmatter}


\section{Introduction}

Very recently, the TOTEM experiment released the following values at $\sqrt{s}$ = 13 TeV of $pp$ total cross section $\sigma^{pp}$ and $\rho^{pp}$ parameter \cite{RRB}
\begin{equation}\label{eq:Totem-data}
\sigma_{tot}^{pp}=110.6\pm 3 \rm{mb}, \qquad \rho^{pp}=0.098 \pm 0.01 
\end{equation}
As it can be seen from (\ref{eq:Totem-data}), the experimental uncertainty of $\rho^{pp}$ is very small.

The value of $\sigma_{tot}$ is in good agreement with the standard best COMPETE prediction \cite{COMPETE} but is in violent disagreement with the  COMPETE prediction for $\rho^{pp}$ (which is much higher than the experimental value). This is the first enigma we have to solve before driving conclusions about the discovery of the Odderon (which is absent in the COMPETE approach). On another si\-de, the experimental value of $\rho^{pp}$ is in perfect agreement with the Avila-Gauron-Nicolescu (AGN) model \cite{AGN}, which includes the Odderon and which predicts a value of 0.105. In fact, the AGN model is the only existing model which correctly predicts $\rho^{pp}$ but it predicts also higher values of $\sigma_{tot}$ than the TOTEM values, a discrepancy which might be connected with the ambiguities in prolonging the amplitudes in the non-forward region. This is the second enigma we have to solve before driving conclusions about the discovery of the Odderon.

We therefore decided to make a careful analysis of the forward proton-proton and antiproton-proton data from very low energies ($\sqrt{s} >$ 5 GeV) till the LHC energies.

But before describing our results, it is important to make a very short review on the Odderon.

The Odderon is defined as a singularity in the complex $j$-plane, located at $j=1$ when $t = 0$ and which contributes to the odd-under-crossing amplitude $F_-$. It was first introduced in 1973 on the theoretical basis of asymptotic theorems \cite{L-N}. The name ``Odderon''  itself was introduced two years later \cite{JLNL}. There is a large variety of possible Odderons, but the most esthetical case is that of the Maximal Odderon (MO), considered in \cite{L-N}. MO leads to the prediction that the difference between particle-particle and particle-antiparticle total cross-sections is not going to 0 at very high energies, contrary to the beliefs of that epoch. It can even lead to a proton-proton total cross-section bigger than antiproton-proton total cross-section, a situation considered, when the Odderon was introduced, as revolutionary, if not heretical. 
In the same year 1973, an important experimental discovery was made at ISR: $\bar pp$ was growing like $\ln^2s$, the maximal behaviour allowed by general principles. This maximal behaviour was first introduced by Heisenberg in 1952, on the basis on geometrical considerations \cite{W.Heis}. A rigorous demonstration was given nine years later by Froissart \cite{F}. The corresponding maximal behaviour of the even-under-crossing amplitude $F_+$ is
\begin{equation}\label{eq:Froissaron}
F_+(s,t=0)\propto s[i\ln^2 s+\pi \ln s]
\end{equation}
The authors of \cite{L-N} established the corresponding maximal behavior of the odd-under-crossing amplitude $F_-$
\begin{equation}\label{eq:Odderon}
F_-(s,t=0)\propto s[i\pi \ln s-\ln^2s]
\end{equation}
leading, via the optical theorem, to the difference of the antiparticle-particle and particle-particle total cross-sections
\begin{equation}\label{eq:delsig}
\Delta \sigma \propto \ln s
\end{equation}
which grows, in absolute value, with energy. However, the sign of $\Delta\sigma$ is not fixed by general principles.

The behaviour (\ref{eq:Froissaron}) is often referred to as ``Froissaron'', while the behaviour (\ref{eq:Odderon}) is termed as ``Maximal Odderon''. At $t=0$, the Froissaron corresponds in the $j$-plane to a triple pole located at $j=1$, while the Maximal Odderon corresponds to a double pole located at $j=1$ \cite{GLN}.
One has to remark that the Maximal Odderon (MO) corresponds to a nice symmetry of the analytic behaviour of $F_+$ and $F_-$ at very high energies:
\begin{equation}\label{eq:symmetry}
\begin{array}{ll}
&\text{Re}F_-(s,0)\sim \text{Im}F_+(s,0)\\ 
&\text{Im}F_-(s,0)\sim \text{Re}F_+(s,0)
\end{array}
\end{equation}
Seven years after the introduction of the concept of the Odderon, it was rediscovered in QCD as a compound state of 3 reggeized gluons \cite{BKP}. A solution corresponding to an intercept exactly equal to 1 was found in 2000 by L. N. Lipatov and collaborators \cite{BLV}.

On the experimental level, from 44 years now, there was just one indication for the presence of the Odderon: the experimental discovery at ISR, in 1985, of a difference between $d\sigma^{\bar pp}/dt$ and $d\sigma^{pp}/dt$ at $\sqrt{s}$= 52.8 GeV in the dip-shoulder region $t = -1.3$ GeV$^2$ \cite{Breaks} . The data were obtained in one week, just before ISR was closed, and therefore the statistical significance of this effect was not very impressive.

The fact that the Odderon was not yet discovered till now is, in fact, not very surprising. The Odderon is about the differences between particle-particle and antiparticle-particle scatterings at high energies. But at high energies we simply do not have both proton-proton and antiproton-proton colliders functioning at the same energy. Moreover, the asymptotic Froissaron and Maximal Odderon contributions appear first, at low energies, as small corrections to Regge-pole physics. We can have a measure of what ``high-energy'' might mean by observing that the minimum in the total cross-sections in the ISR region is approximately 40 mb, a value well described by the familiar Pomeron Regge pole. ISR region is still a low-energy region. In the LHC energy region, the values of total cross-sections attain 110 mb, i. e. the contributions of the high-energy terms is around 70 mb as compared with 40 mb. In the LHC region we really begin to penetrate in the high-energy region. The big hope is the discovery of the Odderon in the LHC region by measuring the phase of the amplitudes, i. e. $\rho $ and $d\sigma/dt$. It is precisely in this context that the TOTEM datum for $\rho $ at $\sqrt{s}$ = 13 TeV is crucial.

We define our Froissaron-Maximal Odderon (FMO) model, valid at $t=0$  as a superposition of the Froissaron, Maximal Odderon and secondary Reggeons contributions. The leading Regge-pole contributions (Pomeron and Odderon Regge poles) are absorbed in the constants present in the Froissaron and Maximal Oddeon contributions.
The amplitudes $F_\pm(s,t)$ are defined to be
\begin{equation}\label{eq;}
F_\pm (z_t, t) = (1/2) (F_{pp}(z_t, t) \pm F_{\bar pp}(z_t, t)) 			
\end{equation}
where  $z_t=\cos\Theta_t$ and $\Theta_t$ is a scattering angle in $t$-channel of the process $pp\to pp$, and  $z_t(t=0)\equiv z= (s-2m^2)/2m^2$. Because we consider here only $t=0$ in what follows we omit for a simplicity the second argument in the amplitudes and write $F(z_t,t=0 )\equiv F(z)$. The amplitudes are normalized so that
\begin{equation}\label{eq:sigdef}
\sigma_{tot}(s) =\text{Im}F(z)/ \sqrt{(s (s-4m^2 )} 
\end{equation}
\begin{equation}\label{eq:rhodef}
\rho(s) = \text{Re}F(z)/\text{Im}F(s). 				
\end{equation}
$m$ in Eq.(\ref{eq:sigdef})  is the mass of the proton.
The amplitude of proton-proton is
\begin{equation}\label{eq:ppampl}
F_{pp}(z) = F^H_+(z) + F^{MO}_-(z) + F^R_+(z) + F^R_-(z)
\end{equation}
and the amplitude of antiproton-proton scattering is			
\begin{equation}\label{eq:papampl}
F_{\bar pp}(z) = F^H_+(z) - F^{MO}_-(z) + F^R_+(z) -F^R_-(z)
\end{equation}

$F^R_+$ in Eqs. (\ref{eq:ppampl}, \ref{eq:papampl}) denotes the contribution of an effective secondary Regge trajectory, whose intercept $\alpha_+(0)$ is located around $j = 1/2$ and is associated with the $f_0(980)$ and $a_0(980)$ particles. Its coupling is denoted as $C^R_{+}$. $F^R_{-}$ in Eqs. (\ref{eq:ppampl},\ref{eq:papampl}) denotes the contribution of an effective secondary Regge trajectory, whose intercept $\alpha_-(0)$ is located around  $j = 1/2$ and is associated with the $\rho (770)$ and $\omega(782)$ particles. Its coupling is denoted as $C^R_{-}$.   
\begin{equation}\label{eq:sec Regge}
F^R_\pm(z)=-\binom{1}{i}
C^R_\pm (-iz)^{\alpha_\pm(0)}
\end{equation}
These contributions become negligible in the TOTEM range of energy.

$F^H_+$ in Eqs. (\ref{eq:ppampl},\ref{eq:papampl})  denote the Froissaron contribution and is parameterized in terms of 3 parameters $H_i (i=1, 2, 3)$, the capital $H$ recalling the name of ``Heisenberg'':
 \begin{equation}\label{eq:Ft=0}
F^H_+(z)=i(s-2m^2)[H_1\ln^2(-iz)+H_2\ln(-iz)+H_3] 
\end{equation}
$F^{MO}_-(z_t)$ in Eqs. (\ref{eq:ppampl},\ref{eq:papampl}) denote the Maximal Odderon contribution and is parameterized in terms of 3 parameters $O_i  (i=1, 2, 3)$ :
 \begin{equation}\label{eq:MOt=0}
F^{MO}_-(z)=(s-2m^2)[O_1\ln^2(-iz)+O_2\ln(-iz)+O_3] 
\end{equation}

The explicit formulae for the observables in terms of our 10 parameters are:
\begin{equation}\label{eq:sigma}
\begin{array}{ll}
&\sigma_{tot}^{pp}(s)=\dfrac{1-2m^2/s}{\sqrt{1-4m^2/s}}\\
&	\times  \biggl \{H_1\ln^2z+(H_2-\pi O_1) \ln z+H_3-H_1\dfrac{\pi^2}{4}-O_2\dfrac{\pi}{2} \\
&+ C^R_+z^{\alpha_+(0)-1}\sin\biggl (\dfrac{\pi}{2} \alpha_+(0)\biggr )-C^R_-z^{\alpha_-(0)-1}\cos\biggl (\dfrac{\pi}{2}  \alpha_-(0)\biggr )\biggr \}\\ \\
\end{array}	
\end{equation}
\begin{equation}\label{eq:sigma-rho}
\begin{array}{ll}
&\sigma_{tot}^{pp}(s)\rho^{pp}(s)=\dfrac{1-2m^2/s}{\sqrt{1-4m^2/s}}\\
&\times  \biggl \{O_1\ln^2z+(O_2+\pi H_1) \ln z+O_3-O_1\dfrac{\pi^2}{4}+H_2\dfrac{\pi}{2} \\
& - C^R_+z^{\alpha_+(0)-1}\cos\biggl (\dfrac{\pi}{2} \alpha_+(0)\biggr )-C^R_-z^{\alpha_-(0)-1}\sin\biggl (\dfrac{\pi}{2}  \alpha_-(0)\biggr )\biggr \}
\end{array}	
\end{equation}
At $s\to \infty$ we have
\begin{equation}\label{eq:deltasigma-asympt}
\Delta\sigma(s)\equiv \sigma_{tot}^{\bar pp}(s)-\sigma_{tot}^{pp}(s)\to 2O_1\pi\ln(s/m^2)
\end{equation}
\begin{equation}\label{eq:deltarho-asympt}
\Delta\rho(s)\equiv \rho^{\bar pp}(s)-\rho^{pp}(s)\to -2\frac{O_1}{H_1}
\end{equation}

\section{Numerical analysis}

 We fitted all the existing 246 data, including the 6 TOTEM data on $\sigma^{pp}_{tot}$ at $\sqrt{s}$ = 2.76, 7, 8 and 13 TeV and the 2 TOTEM data on $\rho^{pp}(s)$ at $\sqrt{s}$ = 8 and 13 TeV  \cite{RRB}, \cite{totem}, but excluding the ATLAS $\sigma^{pp}$ data \cite{atlas}: the ATLAS $\sigma^{pp}$ data (in particular the 8 TeV datum seems doubtful:  96.07$\pm$0.92 mb as compared with the two independent TOTEM data at 8 TeV 102.9$\pm$2.3 mb and 101.5$\pm$2.1 mb) are incompatible with the TOTEM data and their inclusion would obviously  compromise the coherence of the overall data. 
 
 We get a very satisfactory value of $\chi^2$/dof:
 \begin{equation}\label{eq:chi2}
 \chi^2/\text{dof} = 1.0871 						
  \end{equation}
 
Let us note that we refitted the COMPETE model with
all TOTEM points and found that the model without Odderon
contribution is unable to describe the newest data and
moreover $\chi^2/\text{dof}$ is notably increased. We also
considered the non-Maximal Odderon (by putting $O_1=0$) and
have found that such a model lead to a higher value of $\rho^{pp}$  at 13 TeV.

\begin{figure}[h]
 	\centering
 	\includegraphics[width=0.9\linewidth]{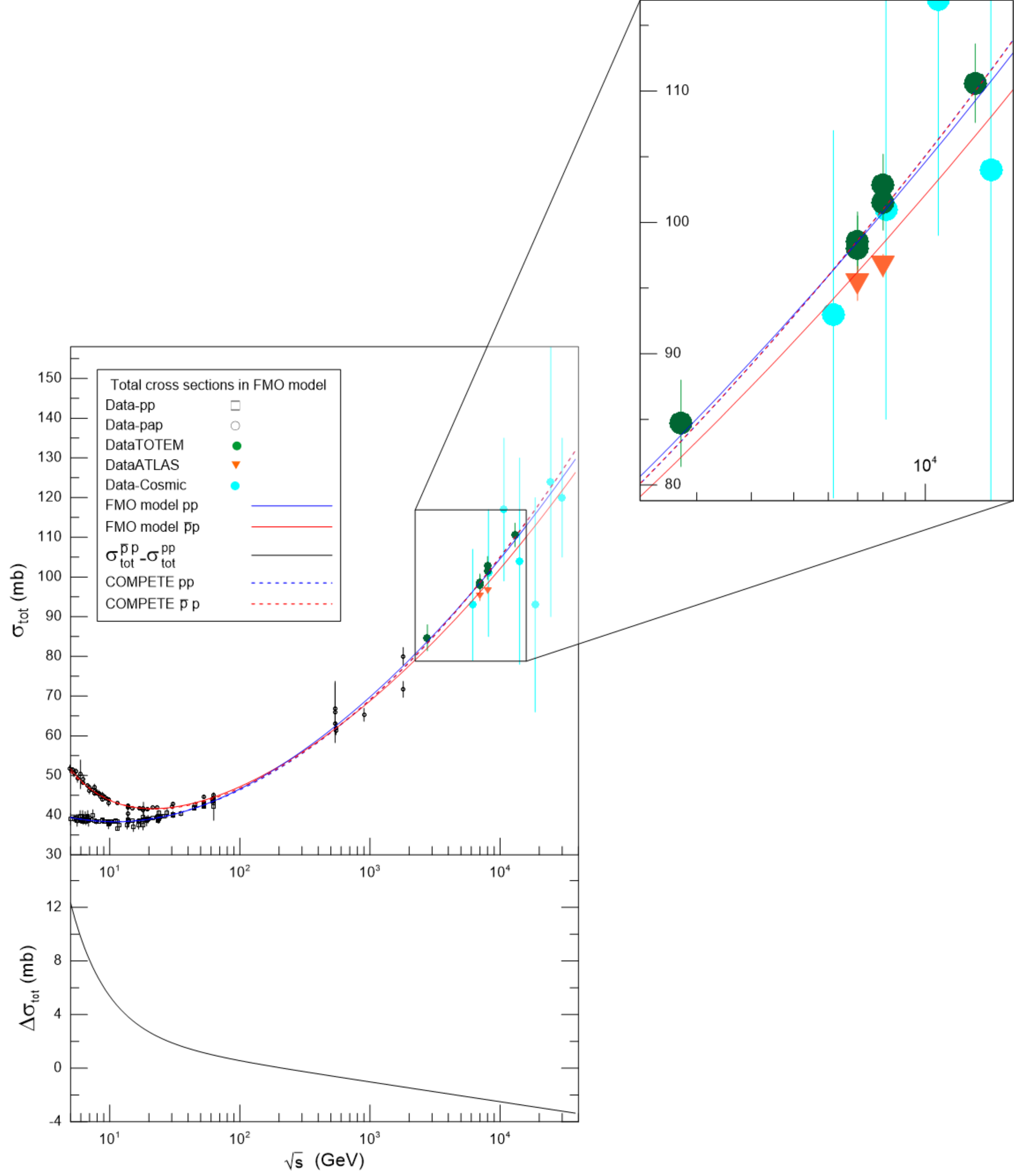}
 	\caption{Total $pp$ and $\bar pp$ cross-sections in FMO model (solid lines). The curves (dashed lines) of the best COMPETE fit \cite{COMPETE} are shown also for a comparison}
 	\label{fig:sigtotfmo}
 \end{figure}

\begin{figure}[h]
 	\centering
 	\includegraphics[width=0.6	\linewidth]{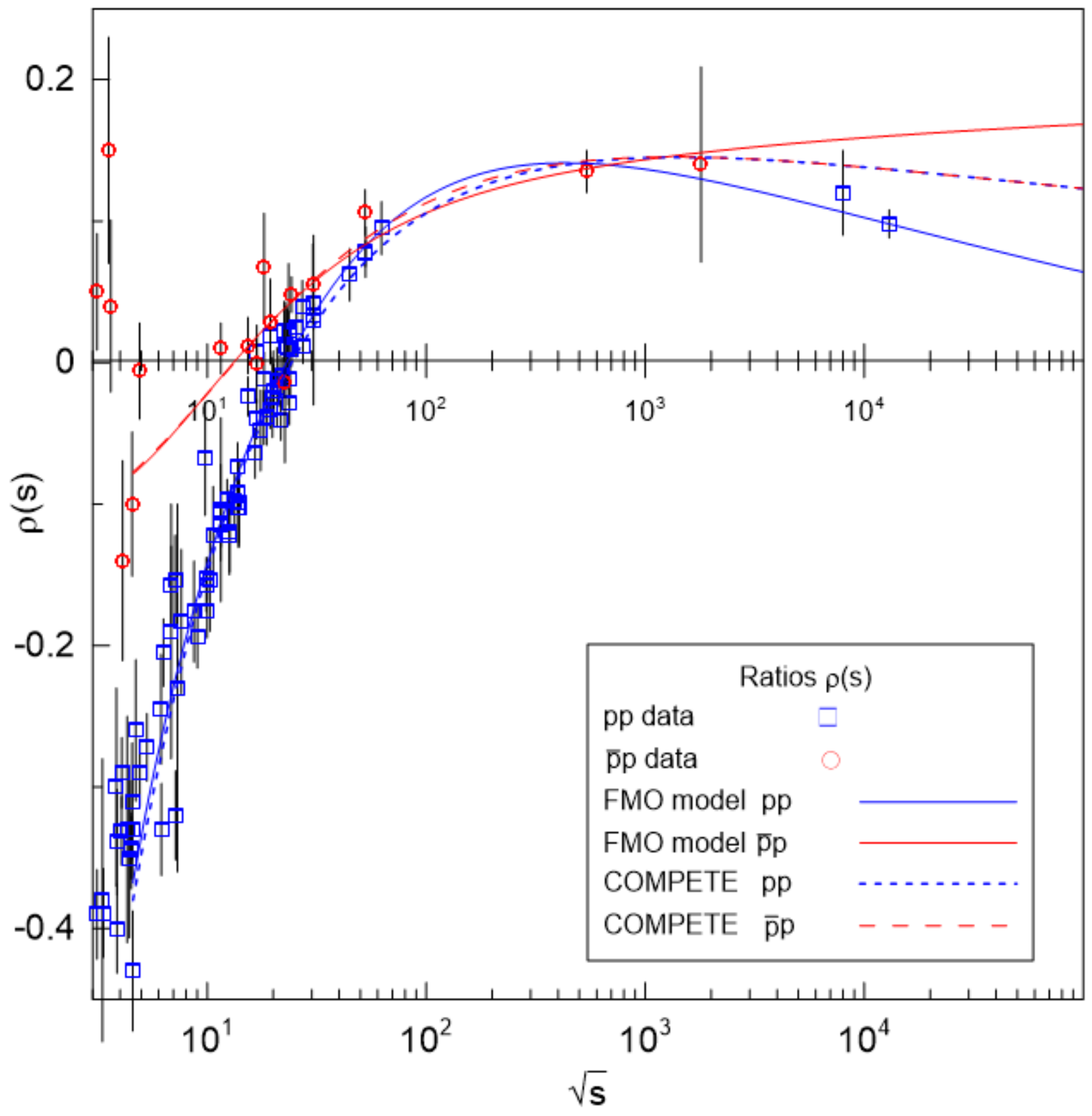}
 	\caption{Ratios of the real to imaginary part of the forward elastic amplitude in FMO model (solid lines). The curves (dashed lines) of the best COMPETE fit \cite{COMPETE} are shown also for a comparison}
 	\label{fig:rho-fmo}
 \end{figure}

 The values of the 10 parameters of the FMO model are shown in the
 Table 1. Table \ref{tab:fmopartials} shows the quality of all data description.
 It can be seen that the values of $H_i$ and $O_i$ are different from the corresponding AGN values. This confirms the fact the discrepancy between good value for $\rho^{pp}$ but higher values for $\sigma^{pp}$ is really connected with the ambiguities in prolonging the amplitudes in the non-forward region.
 In Table 3
 we show all the TOTEM data as compared with the FMO respective values.
 The quality of our fit can be also seen from Figs.\ref{fig:sigtotfmo} and \ref{fig:rho-fmo}.
 In Fig.\ref{fig:sigtotfmo} we show $\sigma^{pp}(s), \Delta \sigma (s)$ and also a magnification of the TOTEM region of energy. 
 
  One can see from Fig.\ref{fig:sigtotfmo} that the best COMPETE curve mimics the values of the FMO model in the TOTEM region of energy. It even crosses the FMO curve at a value of energy around 6 -  8 TeV. This can be easily understood by comparing the explicit formula (\ref{eq:sigma}) with the magic COMPETE formula which can be written at the LHC energies as following 
 \begin{equation}\label{eq:compete}
 \sigma^{pp} (s) = C \ln^2(s/s_0) + P					
 \end{equation}
 where $P$ = 35.5 mb, $C$ = 0.307 mb and the scale factor $s_0$ is 29 GeV$^2$. This coincidence is just a numerical accident of recombination of parameters, without physical significance.
 Let us also note from Fig. \ref{fig:sigtotfmo} that $\Delta \sigma (s)$ has certainly negative values (as a result of the sign of the $O_1$ parameter - see Table 1) 
   but the absolute value is quite small in the TOTEM region (few mb).
 

 In Fig.\ref{fig:rho-fmo} one can see the big disagreement between the $\rho^{pp}$ value at 13 TeV and the best COMPETE prediction. In fact, the small value of $\rho^{pp}$ at 13 TeV confirms the tendency of a decreasing $\rho^{pp}$ already visible at 8 TeV: $\rho^{pp}= 0.12 \pm 0.03.$  One can see also that, in the FMO model, $\rho^{pp}$ and $\rho^{\bar pp}$ cross each other two times: once at around 60 GeV and a second time around 600 GeV. Therefore, in the TOTEM region of energy $\rho^{\bar pp}$ is predicted to be higher (as a result of the sign of the $O_1$ parameter - see Eq. (\ref{eq:deltarho-asympt}) and Table 1) than $\rho^{pp}$ by a substantial amount. 
 
 \renewcommand{\arraystretch}{1.6}
 
 \begin{table}\label{tab:parameters}
 	\centering
 	\begin{tabular}{llp{0.6in}}
 		\hline
 		Parameter     & Value    & Error \\ \hline
 		$H_1$ (mb)    &  0.24964  & $^{+0.02452}_{-0.03058}$\\
 		$H_2$ (mb)    & -0.31854 & $^{+0.86147}_{-0.63729}$\\
 		$H_3$ (mb)    & 30.012   & $^{+4.304}_{-6.363}$      \\
 		$O_1$ (mb)    & -0.05098 &   $^{+0.01050}_{-0.00989}$      \\
 		$O_2$ (mb)    &   1.0240           &   $^{+0.1883}_{-0.2097}$      \\
 		$O_3$ (mb)    &   -4.9110           &   $^{+1.0319}_{-0.8852}$      \\
 		$\alpha_+(0)$ &  0.62957           &   $^{+0.05042}_{-0.04803}$      \\
 		$C^R_+$ (mb)    &  47.292             &   $^{+3.862}_{-2.879}$      \\
 		$\alpha_-(0)$ & 0.26530             &   $^{+0.06727}_{-0.07050}$       \\
 		$C^R_-$ (mb)    &   36.113             &   $^{+5.177}_{-3.790}$      \\
 		\hline
 	\end{tabular}
 	\caption{The values of parameters of FMO model}
 \end{table}

 \renewcommand{\arraystretch}{1.3}
  \begin{table}[h]
 	\centering
 	\begin{tabular}{ccc}
 		\hline 
 		Observable& Number of points & $\chi^2/N_p$	\\ 
 		\hline 
 		$\sigma_{tot}^{pp}$& 110 & 0.8486	\\ 
 		$\sigma_{tot}^{\bar pp}$& 59 & 0.8662\\ 
 		$\rho^{pp}$& 66 & 1.6088\\ 
 		$\rho^{\bar pp}$& 11 & 0.5468 \\ 
 		\multicolumn{2}{c}{ $\chi^2/\text{dof}$} & 1.0871 \\
 		\hline 
 	\end{tabular} 
 	\label{tab:fmopartials}
 	\caption{Number of experimental points $N_p$ and $\chi^2/N_p$ for $\sigma_{tot}$ and $\rho$ in the fit with FMO model}
 \end{table}

\renewcommand{\arraystretch}{1.2}
\begin{table}[h]\label{tab:3}
		{\small
	\begin{tabular}{ccccc}
		\hline 
		$\sqrt{s}$ (TeV) & \multicolumn{2}{c}{$\sigma^{pp}_{tot}$ (mb)}&\multicolumn{2}{c}{$\rho^{pp}$}\\ 
		\hline 
		& TOTEM & FMO & TOTEM &FMO \\ 
		2.76 & 84.7$\pm$ 3.3 & 83.66 & - & 0.123 \\ 
		\multirow{2}*{7} & 98.6$\pm$ 2.2 &\multirow{2}*{98.76} &\multirow{2}*{ -}&\multirow{2}*{0.109}\\ 
		& 98.0$\pm$ 2.5 &  &  &  \\ 
		\multirow{2}*{8}& 101.5$\pm$ 2.1 & \multirow{2}*{101.09}&\multirow{2}*{0.12 $\pm$ 0.03}&\multirow{2}*{ 0.106} \\ 
		& 102.9$\pm$ 2.3 &  &  &  \\ 
		13 & 110.6$\pm$ 3 & 109.92 & 0.098$\pm$0.01&0.0976 \\ 
		\hline 
	\end{tabular} 
}
	\caption{TOTEM data and the best fit values in FMO model for the $\sigma_{tot}$ and $\rho$}
\end{table}


\section{The theoretical status of the FMO approach}
Let us also analyze the theoretical status of the FMO approach.

Both the Froissaron and the Maximal Odderon were introduced on the basis of general principles. 

The Maximal Odderon is a particular case of the Odderon. When the Odderon was introduced \cite{L-N,JLNL}, it was considered as a big surprise and generated a lot of polemics \cite{polemics}. But now not only it was rediscovered in QCD but also in approximations schemes for QCD : the Colour Glass Condensate (CGC) approach  \cite{GCC} and in the dipole picture \cite{KSW}. 

In the past, several objections were formulated against the Froissaron and the Maximal Odderon:
\begin{enumerate}
	\item 	It was shown rigorously in 1967 by Lukaszuk and Martin \cite{L-M} that the constant in front of $\ln^2s$ in $\sigma_{tot}$ is bounded by  $\pi/m_{\pi}^2$ which is approximately 60 mb, a value much bigger than the phenomenological value (for example, in our case, 0.25 mb) and it was hastily concluded that the Froissaron has nothing to do with the saturation of the Froissart bound \cite{F}. But this argument is wrong, because the Lukaszuk-Martin bound takes into account only elastic unitarity and can probably be much improved. The small value of the phenomenological constant does not mean that the growth is not maximal in the functional sense.  As was explained for example in Ref. \cite{L-N}, the asymptotic $\ln^2s$ behaviour is perfectly compatible with any value of the constant between 0 and $\pi/m_\pi ^2$.
	\item In 1997, Andr\'{e} Martin formulated a theorem stating that if, in the strip $−T < t < 0,\quad  s > s_M$ (where $T$ is arbitrarily small and $s_M$ is arbitrarily large), the difference of the imaginary parts of the amplitudes $AB \to AB$ and $A\bar B \to A\bar B$ has a constant sign, and if $d\sigma/dt\to 0$ for both reactions, the difference of the total cross-sections $\Delta \sigma$ does not tend to infinity, and therefore the Maximal Odderon is excluded \cite{M}. However, the particular assumptions made in this theorem exclude an entire class of Odderons (and, between them, the Maximal Odderon). Therefore this theorem is tautological and has no general validity.
	
	\item 	In 2009 Troshin \cite{T}, based upon the principle of maximum strength of strong interactions formulated in 1960-1962 by Chew and Frautschi \cite{ChF} and containing the requirement of saturating unitarity at infinite energies, concluded that the Maximal Odderon is excluded because it does not saturate unitarity and that at asymptotic energies $\rho(s)\to 0$ (result which exclude, in fact, a large class of Odderons).  The principle of maximum strength of strong interactions -- {\it strong interactions are as strong as possible} -- is, of course, very interesting and but was formulated in a time when only Regge poles were believed to exist as $j$-plane singularities and when everybody was convinced that the total cross sections are constant at infinite energies. But, in the presence of other singularities than Regge poles, the condition of saturating unitarity is simply no more a necessary condition: it is sufficient to satisfy unitarity. In collaboration with Lukaszuk and Gauron, we showed the consistency of the Maximal Odderon approach with the QFT constraints \cite{GLN-92}. Namely, we presented a class of amplitudes with the Maximal Odderon type of asymptotics and simultaneously consistent with s-channel unitarity, fixed-$t$ analyticity and the absence of $j=1$ massless state in the $t$-channel. In fact, the FMO approach embodies a new form of the principle of maximum strength of strong interactions: both the even and the odd-under-crossing amplitudes saturate the asymptotic bounds.
	
	\item In 2009, Avsar, Hatta and Matsuo \cite{AHM}, based on the AdS/CFT correspondence, showed that, due to the warp factor of AdS$_5$, the single Odderon exchange gives a larger total cross section in baryon -- baryon collisions than in baryon -- antibaryon collisions at asymptotically high energies, in agreement with our FMO prediction.
\end{enumerate}

We would like to stress that there is no general argument against the FMO approach.

\section{Conclusion} 

Our present study shows that the new TOTEM datum ρ$\rho^{pp} = 0.098\pm 0.01$  can be considered as the first experimental discovery of the Odderon, namely in its maximal form. The present article, in its preprint form, had already a follow-up in the paper by Khoze, Martin and Ryskin, which confirmed the fact that the Odderon must be present in order to explain the new TOTEM datum ρ at 13 TeV \cite{KMR}. The next task is to extend the FMO model for $t$ different from 0. The precise TOTEM $d\sigma/dt$ data at 13 TeV are highly expected to bring other important lights on Odderon effects: the dip mechanism in $d\sigma/dt$ is intimately connected with the predominance of the odd-under-crossing amplitude (the Maximal Odderon amplitude) at high energies.

\medskip
{\bf Acknowledgements.} The authors thank Yoshitaka Hatta, Marcio Menon, Vladimir Petrov, Carlo Ewerz, Mateus Broilo and Wlodek Guryn for a careful reading of our manuscript.

\medskip

\end{document}